\documentstyle[12pt,epsf,amsfonts]{article}
\title{Radiation reaction and renormalization\\
via conservation laws of the Poincar\'e group }
\author{Yurij YAREMKO}
\date{Institute for Condensed Matter Physics of NASU,\\
1 Svientsitskii St., UA--79011 Lviv, Ukraine\\
E-mail: yar\symbol{"40}ph.icmp.lviv.ua}
\begin{document}

\maketitle

\begin{abstract}
We consider the self-action problem in classical electrodynamics of a 
point-like charge arbitrarily moving in flat space-time of four or six 
dimensions. A consistent regularization procedure is proposed which exploits 
the symmetry properties of the theory. The energy-momentum and angular 
momentum balance equations allow us to derive the radiation reaction forces 
in both 4D and 6D. It is shown that a point-like source in 6D possesses an 
internal angular momentum with magnitude which is proportional to the square 
of acceleration. 6D action functional contains, apart from usual "bare" 
mass, an additional renormalization constant which corresponds to the 
curvature of the world line (i.e. to the magnitude of internal angular 
momentum of "bare" particle). It is demonstrated that the {\it 
Poincar\'e-invariant} six-dimensional electrodynamics is renormalizable 
theory.
\end{abstract}



\section{Introduction}
Recently \cite{Gl,KLS}, there has been considerable interest in 
renormalization procedure in classical electrodynamics of a point particle 
moving in flat space-time of arbitrary dimensions. The main task is to 
derive the analogue of the well-known Lorentz-Dirac equation \cite{Dir}. 
Following scheme proposed by Dirac in his classical paper \cite{Dir}, 
Gal'tsov in \cite{Gl} decomposes three vector potential of a 
point-like charge $A_{ret}^\mu=A_S^\mu+A_R^\mu$. The first term, $A_S^\mu$, 
is the half-sum of the retarded, $A_{ret}^\mu$, and the advanced, 
$A_{adv}^\mu$, solutions of the D'Alembert equation with point-like source. 
Since $A_S^\mu$ is singular in the immediate vicinity of the world line, 
the subscript "S" stands for "singular" as well as "symmetric". Because 
$A_S^\mu$ is just singular as $A_{ret}^\mu$, removing it from the retarded 
solution gives the potential $A_R^\mu=1/2(A_{ret}^\mu-A_{adv}^\mu)$ that 
well behaves near the world line. Since $A_R^\mu$ satisfies the homogeneous
wave equation, it can be interpreted as a free radiation field. Hence
the subscript "R" stands for "regular" as well as "radiative".

In classical electrodynamics of a point-like charge arbitrarily moving in 
flat space-time of four dimensions the singular part gives a divergent 
self-energy while the regular one leads to standard radiation reaction 
force. The unphysical "bare" mass involved in action integral absorbs the 
divergent self-energy of a point charge within the renormalization 
procedure and then becomes the observable finite rest mass of the particle 
\cite{Dir,Teit,Gl}. In six dimensions the Coulomb potential of a charge 
scales as $|{\bf r}|^{-3}$ \cite{Gl,KosPr}. Inevitable infinities arising in 
higher-dimensional electrodynamics are stronger than in four dimensions. 
For this reason Gal'tsov \cite{Gl} claims, that "in even dimensions higher 
than four, divergences cannot be removed by the mass renormalization". 

To make classical electrodynamics in six dimensions a renormalizable theory,
in \cite{Kos} the six-dimensional analogue of the relativistic particle 
with rigidity \cite{Pl,Pv,Nest} is substituted for the structureless point 
charge whose Lagrangian is proportional to worldline length. New
Lagrangian involves, apart from usual "bare mass", an additional 
regularization constant which absorbs one extra divergent term. In 
\cite{KLS} the procedure of regularization in any dimensions is elaborated
by using of functional analysis. To get a renormalizable theory for even 
dimension $D>4$, the original Lagrangian for free relativistic particle 
with "bare" mass is modified by introducing $D/2-2$ extra (higher 
derivative) terms. So, in six dimensions the whole renormalization 
procedure involves two arbitrary constants. Since there is three types of 
divergences in six dimensions, Gal'tsov doubts whether " all divergences 
arising in higher-dimensional electrodynamics can be absorbed in this way" 
\cite{Gl}. Introduction of higher derivatives in the particle action term 
seems for him not reasonable enough since it "drastically changes the 
initial theory". 

Contrary to \cite{Gl,KLS} where the authors deal with equations of motion,
Kosyakov \cite{Kos} calculates the flux of energy-momentum and derive the 
radiation-reaction force by adding of appropriate Schott term. In the 
present paper we consider also the conserved quantities corresponding to the 
invariance of the theory under proper homogeneous Lorentz transformations. 
They gives an additional information which clarifies the essence of 
renormalization procedure. In such a way we reformulate the problem of 
renormalizability within the problem of Poincar\'e invariance of a closed 
particle plus field system. The conservation laws are an immovable fulcrum 
about which tips the balance of truth regarding renormalization and 
radiation reaction. Either nonrenormalizable theory or renormalizable one 
should be compatible with the Poincar\'e symmetry.

\section{Preliminaries}
The standard variational principle is formulated \cite{Gl,KLS,Kos} for a 
composite system of point-like charged particle and its own electromagnetic 
field:
\begin{equation}\label{Itot}
I_{\mbox{\scriptsize total}} = I_{\mbox{\scriptsize part}} + 
I_{\mbox{\scriptsize int}} + I_{\mbox{\scriptsize field}}.
\end{equation}
Here
\begin{equation}\label{If}
I_{\mbox{\scriptsize 
field}}=-\frac{1}{4\Omega_{D-2}}\int d^DyF^{\mu\nu}F_{\mu\nu}\,,
\quad
I_{\mbox{\scriptsize part}}=-m\int d\tau \sqrt{-\dot{z}^2},
\end{equation}
and the interaction term given by
\begin{equation} \label{IA}
I_{\mbox{\scriptsize int}} = e\int d\tau A_\mu 
{\dot z}{}^\mu .
\end{equation}
The particle's world line $\zeta:{\Bbb R}\to {\Bbb M}_D$ is described 
by the functions $z^\alpha(\tau)$ which give the particle's coordinates as 
function of proper time $\tau$; ${\dot 
z}^\alpha(\tau)=dz^\alpha(\tau)/d\tau$. By $\Omega_{D-2}$
we denote the area of a $(D-2)$-dimensional sphere of unit radius:
\begin{equation}
\Omega_{D-2}=2\frac{\pi^{(D-1)/2}}{\Gamma(\frac{D-1}{2})}.
\end{equation}

The action (\ref{Itot}) is invariant under infinitesimal 
transformation (translations and rotations) which constitute the Poincar\'e 
group. According to Noether's theorem, these symmetry properties can be used 
for the derivation of conservation laws, i.e. those quantities that do not 
change with time.

Strictly speaking, the action integral (\ref{Itot}) may be used to derive 
trajectories of the test particles, when the field is given {\em a 
priory}. It may also be used to derive $D-$dimensional Maxwell equations, 
if the particle trajectories are given {\em a priory}. Simultaneous 
variation with respect to both field and particle variables is incompatible
since the Lorentz force will always be ill defined in the immediate 
vicinity of the particle's world line. 

Our consideration is founded on the field (\ref{If}) and the interaction 
(\ref{IA}) terms of the action (\ref{Itot}). They constitute the action 
functional which governs the propagation of the electromagnetic field 
produced by a moving charge (i.e. the Maxwell equations with point-like 
source): 
\begin{equation} \label{Maxwell}
\Box A^\alpha(y)=-\Omega_{D-2}j^\alpha(y) .
\end{equation}
Li\'enard-Wiechert fields are the solutions of Maxwell 
equations with point-like sources. These "fields" do not have degrees of 
their own: they are functionals of particle paths. One can substitute 
these {\it direct particle fields} \cite{HN} in the conservation laws to 
rewrite them in terms of particle variables. 

The components of energy-momentum carried by the electromagnetic field are 
\cite{Rohr,Kos}
\begin{equation} \label{pem}
p_{\mbox{\scriptsize em}}^\nu (\tau)=P\int_{\Sigma}
d\sigma_\mu T^{\mu\nu} ,
\end{equation}
where $d\sigma_\mu$ is the vectorial surface element on an arbitrary 
space-like hypersurface $\Sigma$. The components of the electromagnetic 
field's stress-energy tensor 
\begin{equation}\label{T}
\Omega_{D-2}T^{\mu\nu} = F^{\mu\lambda}F^\nu{}_\lambda - 
1/4\eta^{\mu\nu} F^{\kappa\lambda}F_{\kappa\lambda}
\end{equation}
have a singularity on a particle trajectory. In 
eq.(\ref{pem}) capital letter $P$ denotes the principal value of the 
singular integral, defined by removing from $\Sigma$ an 
$\varepsilon$-sphere around the particle and then passing to the limit 
$\varepsilon\to 0$.

The angular momentum tensor of the electromagnetic field is written as
\cite{Rohr,LV}
\begin{equation} \label{Mem}
M_{\mbox{\scriptsize em}}^{\mu\nu}(\tau)=P\int_{\Sigma}
d\sigma_\alpha\left(y^\mu T^{\alpha\nu}-y^\nu T^{\alpha\mu}\right) .
\end{equation}

Conservation of the space part $M_{\mbox{\scriptsize em}}^{ij}$ of the 
tensor $M_{\mbox{\scriptsize em}}^{\mu\nu}$ is due to invariance under 
space rotations. Conservation of the mixed space-time components, 
$M_{\mbox{\scriptsize em}}^{0i}$, takes place due to invariance under 
Lorentz transformations.

\section{Coordinate system}

Using the retarded Green function \cite[eq.(3.4)]{Gl} associated 
with the D'Alembert operator $\Box$ and the charge-current density vector 
$e\int d\tau u^\alpha(u)\delta(y-z(u))$ we construct the retarded 
Li\'enard-Wiechert potential in even dimensions:
\begin{equation} \label{Am}
A_\mu=\left(\frac{1}{2\pi}\frac{1}{r}\frac{d}{d 
u}\right)^{(D-4)/2}e\frac{u_\mu(u)}{r}.
\end{equation}

\begin{figure}
\begin{center}
\epsfxsize=7cm
\epsfclipon
\epsffile{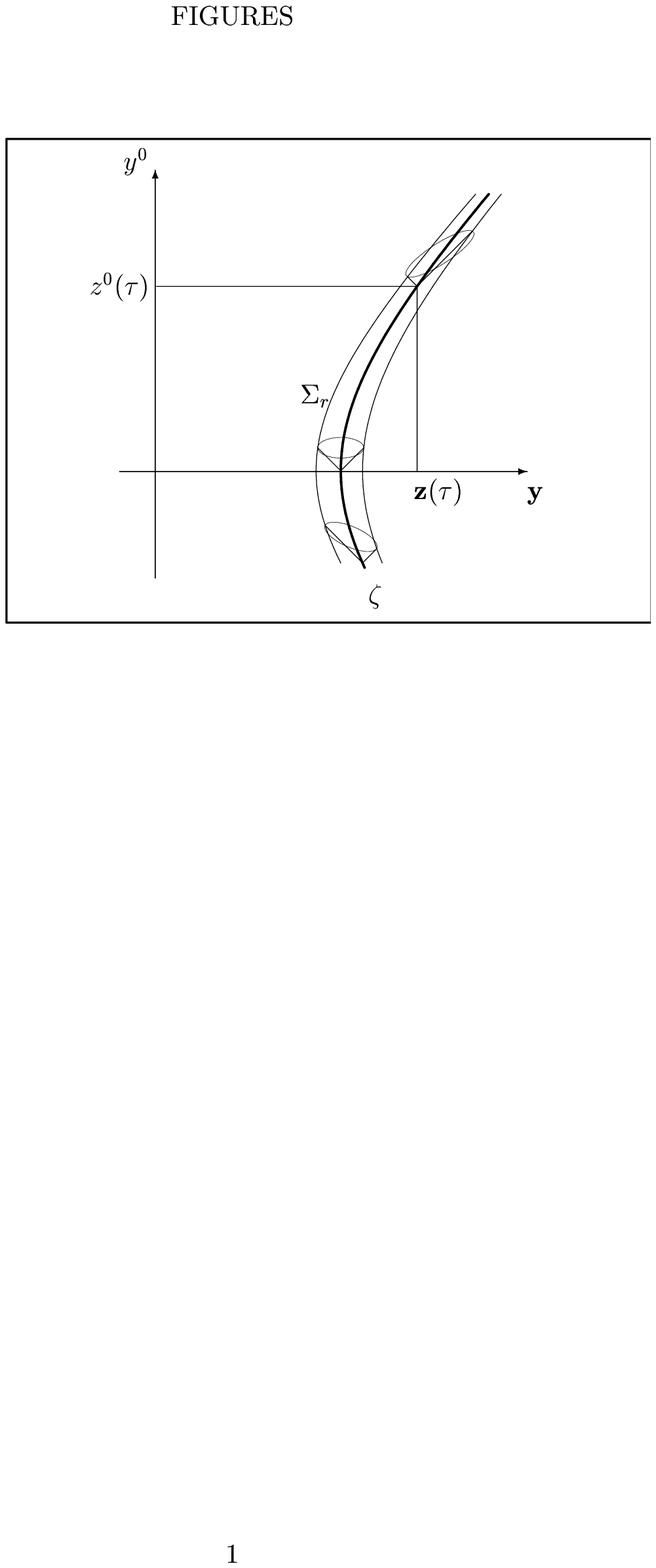}
\end{center}
\caption{Integration region considered in the evaluation of the bound and 
emitted conserved quantities produced by all points of the world line up to 
the end point {\sf E}. Retarded spheres $S(z(u),r), u\in]-\infty,\tau],$ 
of constant radii $r$ constitute a thin world tube $\Sigma_r$ enclosing the 
world line $\zeta$. The sphere $S(z(u),r)$ is the intersection of the future 
light cone with vertex at point $z^\mu(u)\in\zeta$ and $r-$shifted hyperplane
$\Sigma(z(u),r)$ which is orthogonal to particle's velocity $u^\mu(u)$.}
\label{tube}
\end{figure}

An appropriate coordinate system for flat space-time is a very important 
for the volume integration (\ref{pem}) and (\ref{Mem}). 
We calculate how much electromagnetic field momentum and angular momentum 
flow across a world tube of constant radius $r$ enclosing the world line 
$\zeta$ (Bhabha tube \cite{Bh}, see Fig.\ref{tube}). A world tube  is a 
disjoint union of (retarded) spheres of constant radii $r$ centered on a 
world line $\zeta:{\Bbb R}\to {\Bbb M}_D$ of the particle. The sphere 
$S(z(u),r)$ is the intersection of future light cone, generated by null rays 
emanating from $z(u)\in\zeta$ in all possible directions, 
\begin{equation}
C(z(u))=\{y\in {\Bbb M}_D: 
(y^0-z^0(u))^2=\sum_i(y^i-z^i(u))^2,y^0-z^0(u)>0\}
\end{equation}
and tilted hyperplane $\Sigma(z(u),r)$:
\begin{equation} \label{Ksi}
\Sigma(z(u),r)=\{y\in {\Bbb M}_6: 
u_\alpha(u)(y^\alpha -z^\alpha(u)-u^\alpha(u)r)=0\}.
\end{equation}
Points on a sphere are distinguished by spherical polar angles
involved in the space components $n^{i'}$ of the null vector 
$n^{\alpha'}=(1,n^{i'})$, namely
$(\cos\phi\sin\vartheta,\sin\phi\sin\vartheta,\cos\vartheta)$ in four 
dimensions, or $(\cos\varphi\sin\vartheta_1\sin\vartheta_2\sin\vartheta_3,
\sin\varphi\sin\vartheta_1\sin\vartheta_2\sin\vartheta_3,
\cos\vartheta_1\sin\vartheta_2\cdot$\linebreak $\sin\vartheta_3,
\cos\vartheta_2\sin\vartheta_3,
\cos\vartheta_3)$ in six dimensions.

\begin{figure}
\begin{center}
\epsfxsize=8cm
\epsfclipon
\epsffile{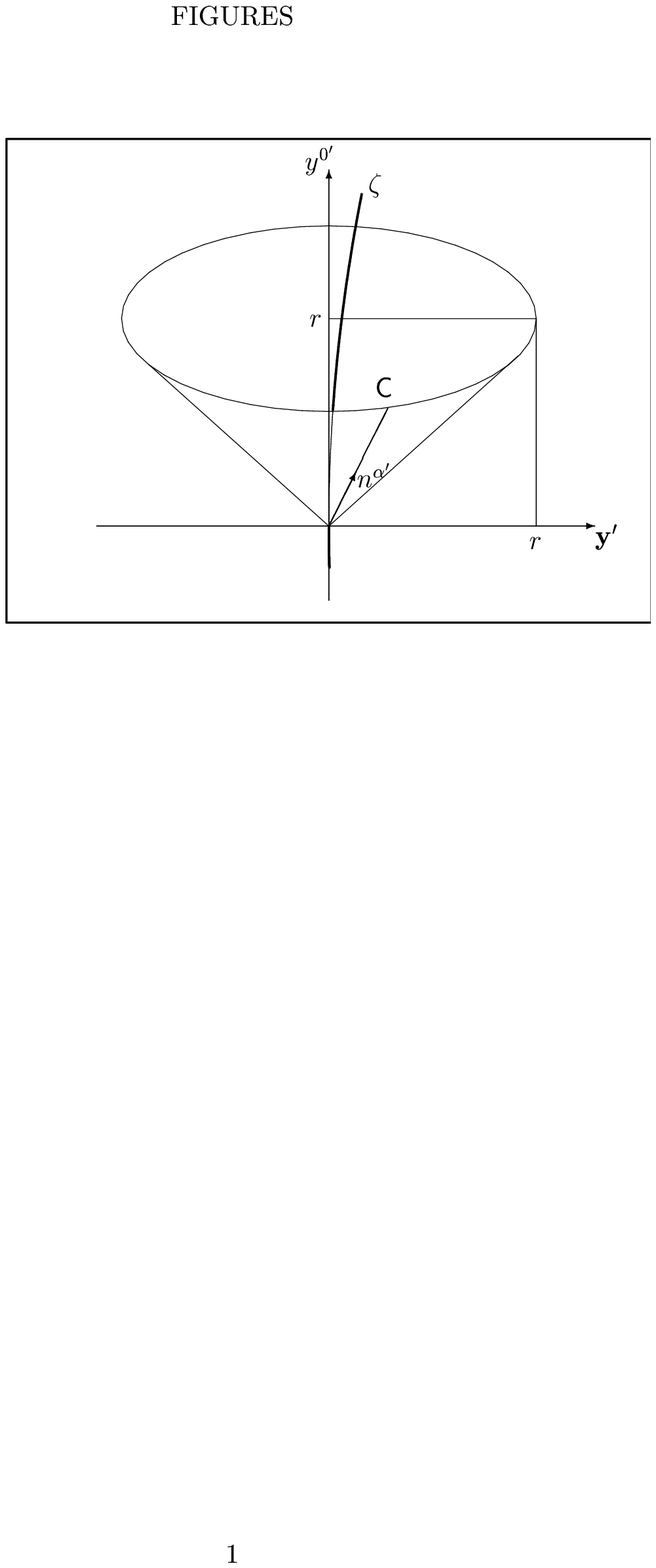}
\end{center}
\caption{In MCLF the retarded distance is the distance between any point on 
spherical light front $S(0,r)=\{y\in {\Bbb M}_6: 
(y^{0'})^2=\sum_i(y^{i'})^2,y^{0'}=r>0\}$ and the particle. The charge is 
placed in the coordinate origin; it is momentarily at the rest.
The point {\sf C}$\in S(0,r)$ is linked to the coordinate origin by a null 
ray characterized by the angles $\vartheta^A$ specifying its direction on 
the cone. The vector with components $n^{\alpha'}$ is tangent to this null 
ray.
}\label{cone}
\end{figure}

To understand the situation more thoroughly, we pass to particle's 
momentarily comoving Lorentz frame (MCLF) where the particle is momentarily 
at the rest at the retarded instant $u$ (see Fig.\ref{cone}). The charge is 
placed in the coordinate origin; the sphere $S(0,r)$ is the intersection of 
the future light cone with vertex at the origin and hyperplane $y^{0'}=r$.

As usual, we call a {\it retarded distance} the distance $r$
between any point on spherical light front $S(0,r)$ and the particle, taken 
in MCLF. In the laboratory frame the points on this sphere have the 
following coordinates:
\begin{eqnarray}\label{y_r}
y^\alpha=&&z^\alpha (u) + r\Lambda^\alpha{}_{\alpha'}(u)n^{\alpha'}
\nonumber\\
=&&z^\alpha (u) + rk^\alpha .
\end{eqnarray} 
The flat space-time ${\Bbb M}_D$ becomes a disjoint union of world 
tubes $\Sigma_r, r>0,$ enclosing the particle trajectory. 

\section{Renormalization and radiation reaction in four dimensions}

\begin{figure}
\begin{center}
\epsfxsize=8cm
\epsfclipon
\epsffile{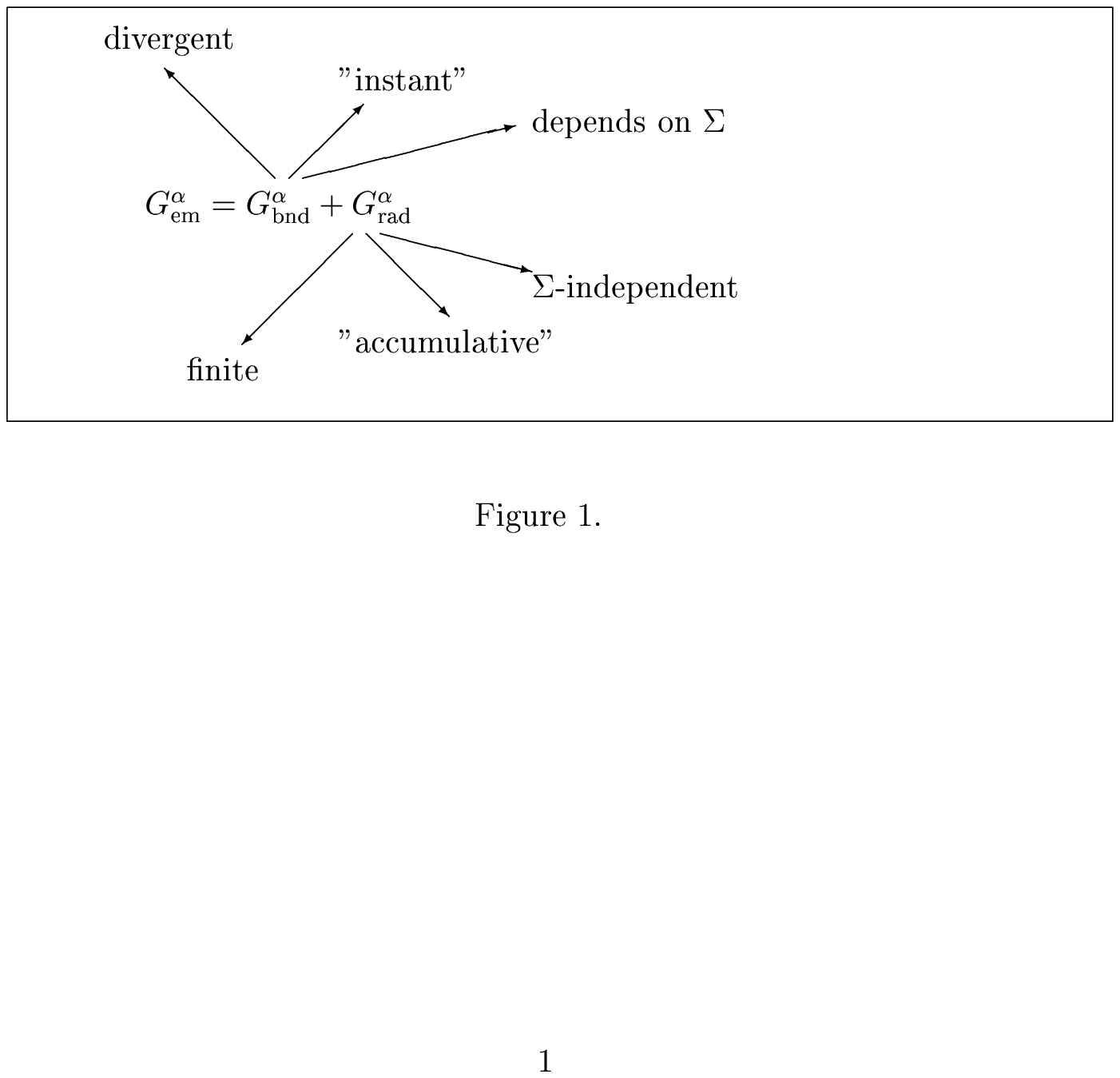}
\end{center}
\caption{
The bound term $G^\alpha_{bnd}$ and the radiative term 
$G^\alpha_{rad}$ constitute Noether quantity $G^\alpha_{em}$ carried by 
electromagnetic field. The former diverges while the latter is finite.
Bound component depends on instant characteristics of charged particles 
while the radiative one is accumulated with time. The form of the bound 
term heavily depends on choosing of an integration surface $\Sigma$ 
while the radiative term does not depend on $\Sigma$. 
}\label{Noe}
\end{figure}

It is straightforward to substitute the components 
$F_{\alpha\beta}=\partial_\alpha A_\beta -\partial_\beta A_\alpha$ into 
eq.(\ref{T}) to calculate the electromagnetic field's stress-energy tensor.
Direct calculations shows \cite{Teit,LV} that either energy-momentum 
(\ref{pem}) or angular momentum (\ref{Mem}) carried by the retarded 
Li\'enard-Wiechert field contains two quite different parts: (i) the bound 
part which is permanently "attached" to the charge and is carried along with 
it; (ii) the radiation part detaches itself from the charge and leads an 
independent existence. The former is divergent while the latter is finite. 
The bound parts depend on the state of particle's motion at the vicinity of 
the observation time only while the radiative parts are accumulated with 
time (see Fig.\ref{Noe}). 
Within regularization procedure the bound terms are 
coupled with energy-momentum and angular momentum of "bare" sources, so 
that already renormalized characteristics $G^\alpha_{part}$ of charged 
particles are proclaimed to be finite. Noether quantities which are 
properly conserved become:
\begin{equation}\label{Noether}
G^\alpha=G^\alpha_{part}+G^\alpha_{rad}.
\end{equation}

On rearrangement, the total four-momentum of our composite particle plus 
field system looks as follows:
\begin{equation}\label{Ptot}
P^\mu=p_{\mbox{\scriptsize part}}^\mu +
\frac23e^2\int_{-\infty}^\tau du (a^\lambda a_\lambda)u^\mu (u),
\end{equation}
where $a^\lambda(u)=d u^\lambda/d u$ is $\lambda$-th component of the 
particle acceleration. Similarly, the total angular momentum is 
\begin{eqnarray}\label{Mtot}
M^{\mu\nu}&=&z^\mu(\tau)p_{\mbox{\scriptsize part}}^\nu 
-z^\nu(\tau)p_{\mbox{\scriptsize part}}^\mu+
\frac23e^2\int_{-\infty}^\tau du (a\cdot 
a)\left[z^\mu(u)u^\nu(u)-z^\nu(u)u^\mu(u)\right]
\nonumber\\
&+&\frac23e^2\int_{-\infty}^\tau du 
\left[u^\mu(u)a^\nu(u)-u^\nu(u)a^\mu(u)\right]
\end{eqnarray}
where dot denotes the scalar product.
Particle four-momentum $p_{\mbox{\scriptsize part}}$ is already renormalized.

Time differentiation of these conserved quantities gives the system of ten 
equations in four variables $p_{\mbox{\scriptsize part}}^\mu$ and their time 
derivatives \cite{Yar}. Having differentiated (\ref{Mtot}) and taking into 
account the differential consequence of eq.(\ref{Ptot})
\begin{equation}\label{dPtot}
{\dot p}_{\mbox{\scriptsize part}}^\mu(\tau)=-
\frac23e^2 (a\cdot a)u^\mu (\tau),
\end{equation}
we arrive at the equality which explains how four-momentum of charged 
particle depends on its velocity and acceleration:
\begin{equation}
u\wedge p_{\mbox{\scriptsize part}}=-\frac23e^2 u\wedge a .
\end{equation}
(The symbol $\wedge$ denotes the wedge product.)
Hence the particle four-momentum contains, apart from the usual velocity 
term, also a contribution from the acceleration when the particle is charged:
\begin{equation} \label{Texp}
p_{\mbox{\scriptsize part}}^\mu = mu^\mu - \frac{2}{3}e^2a^\mu .
\end{equation}
Since $(u\cdot a)=0$, the scalar product of particle four-velocity on the 
first-order time-derivative of particle four-momentum (\ref{dPtot}) is as 
follows:
\begin{equation}\label{pdp}
({\dot p}_{\mbox{\scriptsize part}}\cdot u)=
\frac23e^2 (a\cdot a).
\end{equation}
Similarly, the scalar product of particle acceleration on the 
particle four-momentum (\ref{Texp}) is given by
\begin{equation}\label{pp}
(p_{\mbox{\scriptsize part}}\cdot a)=-
\frac23e^2 (a\cdot a).
\end{equation}
Summing up (\ref{pdp}) and (\ref{pp}) we obtain
\begin{equation}\label{dpdp}
\frac{d}{d\tau}(p_{\mbox{\scriptsize part}}\cdot u)=0.
\end{equation}
Therefore, $m$ is already renormalized rest mass of the particle \cite{KosPr}. 
We are sure that the well-known Teitelboim's expression (\ref{Texp})
for the four-momentum of point-like charged particle \cite{Teit} as well as 
the Lorentz-Dirac equation \cite{Dir} arise from the total energy-momentum 
and angular momentum balance equations.

\section{Radiation reaction in six dimensions}

Taking $D=6$ in (\ref{Am}), one has again that Li\'enard-Wiechert 
potential in six dimensions depends on particle acceleration:
\begin{equation} \label{Amm}
A_\mu=\frac{e}{2\pi}\left[\frac{a_\mu(u)}{r^2}+
\frac{u_\mu(u)}{r^3}\left(1+ra_k\right)\right] .
\end{equation}
Here $a_k=a_\alpha k^\alpha$ is the component of the particle acceleration 
in the direction of $k^\alpha :=\Lambda^\alpha{}_{\alpha'}(u)n^{\alpha'}$.

The direct particle field \cite{HN} is defined in terms of this potential by
$F_{\alpha\beta}=A_{\beta ,\alpha}-A_{\alpha ,\beta}$. Having used the 
differentiation rule \cite[eqs.(2),(3)]{Kos}
\begin{equation}
\frac{\partial u}{\partial y^\mu}=-k_\mu,\qquad  
\frac{\partial r}{\partial y^\mu}=-u_\mu+\left(1+ra_k\right)k_\mu,
\end{equation}  
we obtain
\begin{equation} \label{F}
F=\frac{e}{2\pi}\left(\frac{u\wedge a}{r^3}+V\wedge k\right)
\end{equation}
where
\begin{equation} \label{V}
V_\mu=\frac{3u_\mu}{r^4}+\frac{3(a_\mu +2u_\mu a_k)}{r^3}+\frac{{\dot a}_\mu 
+u_\mu {\dot a}_k + 3a_\mu a_k +3u_\mu a^2_k}{r^2}.
\end{equation}
The overdot means the derivative with respect to retarded time $u$.
Li\'enard-Wiechert field (\ref{F}) coincides with the field obtained in 
\cite[eq.(14)]{Kos} where the "mostly minus" metric signature should be 
replaced by the "mostly plus" one and the overall coefficient $e/2\pi$ 
should be substituted for $1$ (or for $1/3$ in eq.(15) of the E-print 
version corrected by author).

It is straightforward to substitute the components (\ref{F}) into 
eq.(\ref{T}) to calculate the electromagnetic field's stress-energy tensor. 
Following \cite{Kos}, we present $T^{\alpha\beta}$ as a sum of 
radiative and bound components,
\begin{equation} \label{Tr_Tb}
T^{\alpha\beta} = T_{\mbox{\scriptsize rad}}^{\alpha\beta} + 
T_{\mbox{\scriptsize bnd}}^{\alpha\beta}.
\end{equation}
The radiative part scales as $r^{-4}$:
\begin{equation} \label{Tr} 
\frac{8\pi^2}{3}T_{\mbox{\scriptsize rad}}^{\alpha\beta}=
\frac{e^2}{4\pi^2}\frac{k^\alpha k^\beta}{r^4}V_{(-2)}^\mu 
V_\mu^{(-2)}
\end{equation}
where the components $V_\mu^{(-2)}$ of six-vector $V_{(-2)}$ is defined by
eq.(\ref{V}). The others $T_{(-\kappa)}$ constitute the bound part of the 
Maxwell energy-momentum tensor density:
\begin{equation}
T_{\mbox{\scriptsize bnd}}^{\alpha\beta}=T_{(-8)}+T_{(-7)}+T_{(-6)}+T_{(-5)}.
\end{equation}
(Each term has been labelled according to its dependence on the distance $r$.)

According \cite{Kos}, the outward-directed surface element $d\sigma_\mu$ 
of a five-cylinder $r=const$ in ${\Bbb M}_6$ is
\begin{equation} \label{sgm} 
d\sigma_\mu=\left[-u_\mu+(1+ra_k)k_\mu\right]r^4d\Omega_4d u ,
\end{equation}
where 
$d\Omega_4=d\vartheta_1d\vartheta_2d\vartheta_3d\phi\sin\vartheta_1
\sin^2\vartheta_2\sin^3\vartheta_3$ is the element of solid angle in five
dimensions. The angular integration can be handled via the relations
\begin{eqnarray}
\int d\Omega_4=\frac{8\pi^2}{3},\qquad
\int d\Omega_4n^\alpha n^\beta=\frac{8\pi^2}{15}\left(
\eta^{\alpha\beta}+u^\alpha u^\beta\right),\nonumber\\
\int d\Omega_4n^\alpha n^\beta n^\gamma n^\kappa 
=\frac{8\pi^2}{105}\left[
\left(\eta^{\alpha\beta}+u^\alpha u^\beta\right)
\left(\eta^{\gamma\kappa}+u^\gamma u^\kappa\right) 
 \right.
\nonumber\\
+\left.\left(\eta^{\alpha\gamma}+u^\alpha u^\gamma\right)
\left(\eta^{\beta\kappa}+u^\beta 
u^\kappa\right) +
\left(\eta^{\alpha\kappa}+u^\alpha u^\kappa\right)
\left(\eta^{\beta\gamma}+u^\beta u^\gamma\right) \right] .
\end{eqnarray}
The integral of polynomial in odd powers of $n^\alpha:=k^\alpha 
-u^\alpha$ vanishes.

We are now concerned with the integration of (\ref{pem}). 
Volume integration of the bound part of the stress-energy tensor 
over the world tube $\Sigma_r$ of constant radius $r$ reveals 
that the bound energy-momentum is a function of the end points only:
\begin{equation} \label{pbnd}
p_{\mbox{\scriptsize bnd}}^\mu=\frac{e^2}{4\pi^2}
\left[
\frac32\frac{u^\mu(u)}{r^3}+\frac{12}{5}\frac{a^\mu(u)}{r^2} +
2\frac{(a\cdot a)u^\mu(u)}{r}
\right]_{u\to -\infty}^{u=\tau}
\end{equation}
(The matter is that the total (retarded) time derivatives arise from angular 
integration.) If the charged particle is asymptotically free at the remote 
past, we obtain the Coulomb-like self-energy of constant value. The upper 
limit drastically depends on the value of $r$. To evaluate the 
bound part of six-momentum in the neighborhood of the particle we take the 
limit $r\to 0$. If $r$ tends to zero, $p_{\mbox{\scriptsize 
bnd}}^\mu\to\infty$. The divergences are absorbed by individual particle's 
six-momentum within the renormalization procedure based on the Noether 
quantities (see Fig.\ref{Noe} and eq.(\ref{Noether})).
The total energy-momentum of a closed system of an 
arbitrarily moving charge and its electromagnetic field is equal to the sum 
\begin{equation} \label{Pmn}
P^\mu =p_{\rm part}^\mu +p_{\rm rad}^\mu ,
\end{equation} 
where $p_{\rm rad}$ is the radiative part of electromagnetic field's 
energy-momentum which does not depend on $r$ at all:
\begin{equation} \label{prad}
p_{\mbox{\scriptsize rad}}^\mu=\frac{e^2}{4\pi^2}
\int_{-\infty}^\tau du
\left(\frac45(\dot{a}\cdot\dot{a})u^\mu-\frac{6}{35}(a\cdot a)\dot{a}^\mu
+\frac37a^\mu (a\cdot a)^{\bf\cdot}+2(a\cdot a)^2u^\mu\right)
\end{equation}
(We denote $(a\cdot a)^{\bf\cdot}$ the derivative $d(a\cdot a)/du$.) 

Volume integration of the angular momentum tensor density shows, that the 
bound  angular momentum depends on the state of particle's motion at the 
observation instant only:
\begin{equation} \label{Mbnd}
M_{\mbox{\scriptsize bnd}}^{\mu\nu}=\frac{e^2}{4\pi^2}\lim_{r\to 0}\left(
z^\mu P_{\rm bnd}^\nu -z^\nu P_{\rm bnd}^\mu
+\frac{12}{5}\frac{u^\mu a^\nu - u^\nu a^\mu}{r}
\right).
\end{equation}
By symbol $P_{\rm bnd}$ we mean the expression in between the squared 
brackets of eq.(\ref{pbnd}).

It is worth noting that $M_{\mbox{\scriptsize bnd}}^{\mu\nu}$ contains, 
apart from the usual term of type $z\wedge p_{\rm part}$, also an extra term 
which can be interpreted as the "shadow" of internal angular momentum. It 
prompts that the bare "core" possesses a "spin". 

Total angular momentum of our composite particle plus field system
is written as
\begin{equation} \label{Mmn}
M^{\mu\nu}=M_{\rm part}^{\mu\nu} + M_{\mbox{\scriptsize rad}}^{\mu\nu}
\end{equation} 
where $M_{\rm rad}$ is the radiative part (\ref{Mrad}) of electromagnetic 
field's angular momentum which depends on all previous motion of a source:
\begin{eqnarray} \label{Mrad}
M_{\mbox{\scriptsize rad}}^{\mu\nu}=&&\frac{e^2}{4\pi^2}
\left\{
\int_{-\infty}^{\tau} du
\left(
z^\mu P_{\mbox{\scriptsize rad}}^\nu -
z^\nu P_{\mbox{\scriptsize rad}}^\mu 
\right) 
\right.
\nonumber\\ 
&&\left.+
\int_{-\infty}^{\tau} du
\left[
\frac45\left(a^\mu{\dot a}^\nu-a^\nu{\dot a}^\mu\right)
+\frac{64}{35}(a\cdot a)\left(u^\mu a^\nu-u^\nu a^\mu\right)\right]\right\}.
\end{eqnarray}
Here $P_{\mbox{\scriptsize rad}}$ denotes the integrand of 
eq.(\ref{prad}). 

With (\ref{Mbnd}) in mind we assume that already renormalized angular 
momentum tensor of the particle has the form 
\begin{equation} \label{Mprt}
M_{\rm part}^{\mu\nu} =z^\mu p_{\rm part}^\nu-z^\nu p_{\rm part}^\mu +
u^\mu\pi_{\rm part}^\nu -u^\nu\pi_{\rm part}^\mu .
\end{equation} 
In \cite{Pl,Pv,Nest} the extra momentum $\pi_{\rm part}$ is due to 
additional degrees of freedom associated with acceleration involved in 
Lagrangian function for rigid particle.

Our next task is to derive expressions which explains how six-mo\-men\-tum
and angular momentum of charged particle depend on its velocity and 
acceleration etc. Having performed the time differentiation of eq.(\ref{Pmn}) we obtain
the following energy-momentum balance equation:
\begin{equation} \label{pdot}
{\dot p}_{\rm part}^\mu = -\frac{e^2}{4\pi^2}
\left(\frac45(\dot{a}\cdot\dot{a})u^\mu -\frac{6}{35}(a\cdot a)\dot{a}^\mu
+\frac37a^\mu (a\cdot a)^{\bf\cdot}+2(a\cdot a)^2u^\mu\right).
\end{equation} 
(All the particle characteristics are evaluated at the instant of 
observation $\tau$.) Having differentiated (\ref{Mmn}) and taking into 
account (\ref{pdot}) we arrive at the equality which does not contain ${\dot 
p}_{\rm part}$:
\begin{equation} \label{Mdot}
u\wedge\left(p_{\rm part} + {\dot \pi}_{\rm part} +
\frac{e^2}{4\pi^2}\frac{64}{35}(a\cdot a)a\right)
+a\wedge\left[\pi_{\rm part} +\frac{e^2}{4\pi^2}
\frac45\dot{a}\right]=0.
\end{equation}

Scrupulous analysis of consistency of the energy-momentum balance equation
(\ref{pdot}) and angular momentum balance equation (\ref{Mdot}) reveals, 
that six-momentum of charged particle contains two (already renormalized) 
constants \cite[Appendix]{Yar6D}:
\begin{equation} \label{p_ost}
p_{\rm part}^\beta = mu^\beta +\mu\left(-{\dot a}^\beta 
+\frac32(a\cdot a)u^\beta\right)
+\frac{e^2}{4\pi^2}\left[\frac45{\ddot a}^\beta
-\frac85u^\beta (a\cdot a)^{\bf\cdot}-\frac{64}{35}(a\cdot a)a^\beta\right].
\end{equation}
The first, $m$, looks as a rest mass of the charge. But the true rest 
mass is identical to the scalar product of the six-momentum and 
six-velocity \cite{KosPr}. Since the scalar product depends on 
the square of acceleration as well as its time derivative
\begin{equation} 
m_0=-(p_{\rm part}\cdot 
u)=m+\frac{\mu}{2}(a\cdot a)-\frac{e^2}{4\pi^2}\frac25(a\cdot a)^{\bf\cdot} ,
\end{equation}
the renormalization constant $m$ is formal parameter and its 
physical sense is not clear.

The second, $\mu$, is intimately connected with the the internal angular 
momentum $s_{\rm part}:=u\wedge\pi_{\rm part}$ of the particle:
\begin{equation} \label{ss}
s_{\rm part}^{\alpha\beta} = \mu\left(u^\alpha a^\beta -u^\beta 
a^\alpha\right) -\frac{e^2}{4\pi^2}\frac45\left(u^\alpha{\dot a}^\beta
-u^\beta{\dot a}^\alpha \right).
\end{equation} 
But the magnitude of $s_{\rm part}$ is not constant:
\begin{equation} \label{s^2}
s^2=-\frac12s^{\rm part}_{\alpha\beta}s_{\rm part}^{\alpha\beta}
=\mu^2(a\cdot a)+\mu\frac{e^2}{5\pi^2}(a\cdot a)^{\bf\cdot} 
+\frac{e^4}{25\pi^4}\left(
(\dot{a}\cdot\dot{a})+(a\cdot a)^2\right).
\end{equation}
Therefore, this name can not be understand literally.

Having substituted the right-hand side of eq.(\ref{p_ost}) for the 
particle's six-mo\-men\-tum in the energy-mo\-men\-tum balance 
equation (\ref{pdot}),
we derive the Lorentz-Dirac equation of motion of a charged particle under 
the influence of its own electromagnetic field. An external device adds 
covariant external force $F_{ext}$ to the right-hand side of this 
expression.

Expression (\ref{p_ost}) was firstly obtained by Kosyakov in 
\cite[eq.(37)]{Kos} (see eq.(38) in E-print version corrected by author). 
The derivation is based upon consideration of energy-momentum conservation 
only. The author constructs an appropriate Schott term to ensure the 
orthogonality of the radiation reaction force to the particle six-velocity.
The formula define the {\it bound} six-momentum of a point-like charged 
particle which is involved in the energy-momentum balance equation (see 
eq.(40) in E-print version of the paper and eq.(\ref{Pmn}) of the present 
paper where this momentum is denoted as $p_{\rm part}$). The {\it bound} 
momentum differs from one of the particle being "dressed" in 
electromagnetic "fur" (see eq.(42) in E-print version of \cite{Kos}). The 
latter contains also a contribution from the radiated energy-momentum 
(\ref{prad}) carried by electromagnetic field of accelerated particle. The 
six-momentum of "dressed" particle is involved in six-dimensional analogue 
of relativistic generalization of Newton's second law \cite[eq.(38)]{Kos} 
where loss of energy due radiation is taken into account.

We face the problem if {\it two} renormalization constants, $m$ and $\mu$, 
are enough to absorb the {\it three} divergences of our divergent bound 
six-momentum (\ref{pbnd}). The structure of this expression heavily depends 
on the integration surface. Having integrated the electromagnetic field's 
stress-energy tensor (\ref{T}) over hyperplane $\Sigma_t=\{y\in{\Bbb 
M}_6:y^0=t\}$, we obtain
\begin{eqnarray}
p_{\mbox{\scriptsize bnd}}^\mu &=&\int_{\Sigma_t}
d\sigma_0 T_{\mbox{\scriptsize bnd}}^{0\mu}\nonumber\\ 
&=&\frac{e^2}{4\pi^2}
\left[\frac{3}{35}\frac{-12u^0u^\mu +40(u^0)^3u^\mu 
+\eta^{0\mu }\left(-3/2+12(u^0)^2\right)}{(t-u)^3} \right.
\nonumber\\ 
&+&\left.\frac{3}{35}\frac{-5a^\mu + 31a^0u^0u^\mu + 33a^\mu 
(u^0)^2+ 3\eta^{0\mu }a^0}{(t-u)^2}  \right.
\nonumber\\ 
&+&\left.\frac{1}{35}
\frac{37a^0a^\mu + 71(a\cdot a)u^0u^\mu +\eta^{0\mu 
}(a\cdot a)}{t-u}\right]_{u\to -\infty}^{u\to t}.
\end{eqnarray}
The lower limit is equal to zero while the upper one tends to infinity.
The coefficients are quite different from that in eq.(\ref{pbnd}) while the 
number of divergences is still equal to three.

Further we choose the tilted hyperplane $\sigma_\tau=\{y\in{\Bbb M}_6:
u_\mu(\tau)\left(y^\mu-z^\mu(\tau)\right) = 0\}$ which is intimately 
connected with the momentarily comoving Lorentz frame of the charge at the 
observation instant $\tau$ (cf. $r$-shifted hyperplane (\ref{Ksi})). In 
MCLF the particle is momentarily at rest at time $\tau$. To apply our 
previous result we make such Lorentz transformation $\Omega$ that a tilted 
hyperplane $\sigma_\tau$ becomes
$\Sigma_{t'}=\{y\in{\Bbb M}_6 : y^{0'}=t'\}$:
\begin{eqnarray}
p_{\mbox{\scriptsize 
bnd}}^\mu&=&\int_{\sigma_t}d\sigma_\nu T_{\mbox{\scriptsize bnd}}^{\nu\mu}
\nonumber\\
&=&\int_{y^{0'}=t'}d\sigma_{0'}T_{\mbox{\scriptsize bnd}}^{0'\alpha'}
\Omega_{\alpha'}{}^{\mu}\nonumber\\
&=&\Omega^{\mu}{}_{\alpha'}p_{\mbox{\scriptsize bnd}}^{\alpha'}\,.
\end{eqnarray}
It is now straightforward (but tedious) matter to calculate the components 
of bound six-momentum in MCLF. They contains {\it two} divergences only:
\begin{eqnarray}
p_{\mbox{\scriptsize bnd}}^{0'}&=&\frac{e^2}{4\pi^2}
\lim_{u'\to 
t'}\left[\frac32\frac{1}{(t'-u')^3}+2\frac{(a\cdot a)}{t'-u'}\right],
\\
p_{\mbox{\scriptsize bnd}}^{i'}&=&\frac{e^2}{4\pi^2}
\lim_{u'\to t'}\left[-\frac{6}{5}\frac{\dot{a}^{i'}}{t'-u'}\right]
\end{eqnarray}

Since the structure of bound six-momentum is changeable, it would make 
no sense to disrupt the bonds between different powers of small parameter 
$r$ in (\ref{pbnd}). It is sufficient to assume that a charged particle 
possesses its own (already renormalized) six-momentum $p_{\rm part}$ which 
is transformed as an usual six-vector under the Poincar\'e group. 
Careful analysis of the energy-momentum and angular momentum balance 
equations reveals how $p_{\rm part}$ depends on six-velocity, 
six-acceleration etc. In ${\Bbb M}_4$ the solution (\ref{Texp}) contains one 
renormalization parameter: the rest mass $m$. If we deal with 
six-dimensional flat space-time than {\it two} renormalization parameters, 
$m$ and $\mu$, arise (see eq.(\ref{p_ost})).

\section{Conclusions}

We examine whether the renormalizabolity is a necessary condition for 
consistency of the local field theory with fundamental principles such as 
energy-momentum conservation and the conservation of total angular momentum.
A careful analysis with the use of regularization procedure compatible 
with the Poincar\'e symmetry shows, that the usual particle part of initial 
action integral (\ref{Itot}) which is proportional to the worldline 
lenght is inconsistent with $I_{\rm field}$ and $I_{\rm int}$ in six 
dimensions. Indeed, the angular momentum tensor of structureless particle
\begin{equation} 
M_{\rm part}^{\mu\nu} =z^\mu p_{\rm part}^\nu-z^\nu p_{\rm part}^\mu 
\end{equation} 
corresponds to $I_{\rm part}$ given by (\ref{If}). Having analyzed angular 
momentum balance equations one has again
\begin{equation}
u\wedge \left(p_{\rm part}
+\frac{e^2}{4\pi^2}\frac{64}{35}(a\cdot a)a\right) 
+\frac{e^2}{4\pi^2}\frac45 a\wedge\dot a =0
\end{equation}
instead of (\ref{Mdot}). Its solution is a motion with constant velocity 
where $p_{\rm part}^\mu$ do not change. Hence the 
action functional based on the higher-derivative Lagrangian for a "rigid" 
relativistic particle \cite{Pl,Pv} should be substituted for 
$I_{\mbox{\scriptsize part}}$ in (\ref{If}). It involves two 
renormalization constants \cite{KLS,Kos}; it is sufficient to renormalize 
{\em all} the divergences arising in six-dimensional electrodynamics. 

Nonrenormalizable theory contradicts the differential consequences of the 
conserved quantities which arise from the invariance of the system under 
space rotations and Lorentz transformations.

Volume integration of energy-momentum and angular momentum carried by 
electromagnetic field shows, that a charged particle is supplemented with 
the bound electromagnetic "cloud" which has its own (divergent) momentum 
and angular momentum. Corresponding characteristics of "bare" charge absorb 
them within the regularization procedure based on Noether conservation laws
(see Figure \ref{Noe}). The radiative parts of the electromagnetic field's 
energy-momentum and angular momentum detach the charge and lead an 
independent existence. They are involved in energy-momentum and angular 
momentum balance equations which determine how {\it already renormalized} 
particle's momentum and angular momentum depend on its velocity, 
acceleration etc (see eq.(\ref{Texp}) for ${\Bbb M}_4$ and eq.(\ref{p_ost}) 
for ${\Bbb M}_6$).

It is worth noting that in six dimensions a test particle (i.e. point 
charge which itself does not influence the field) is the rigid particle. 
Its momentum is given by the expression (\ref{p_ost}) where $e^2$ tends 
to zero; it is not parallel to six-velocity. The problem of motion of such 
particles in an external electromagnetic field is considered in \cite{Nst}.

\section*{Acknowlegement}
The author would like to thank Professor B.P.Kosyakov,
Professor V.Tre\-tyak,  and Dr.A.Du\-vi\-ryak for 
helpful discussions and critical comments.

\end{document}